\documentclass[11pt,reprint,amsmath,amssymb,
 nofootinbib, aps,prl]{revtex4-2}
 
\usepackage{geometry}
\usepackage{amsmath,amsthm,amssymb, mathtools,mathrsfs}
\usepackage{physics}
\usepackage{amsfonts, latexsym, amssymb}
\usepackage{bbm, dsfont}
\usepackage{graphicx}
\usepackage{float}
\usepackage{url}
\usepackage{xcolor}
\usepackage{hyperref}
\usepackage{tikz}
\usetikzlibrary{decorations.pathmorphing,cd,decorations.markings,calc,fadings}
\hypersetup{
    colorlinks=true,
    linkcolor=blue,
    filecolor=magenta,      
    urlcolor=cyan,
}
\usepackage{enumerate}
\usepackage{bm}
\usepackage{tikz}
\usetikzlibrary{arrows.meta,calc}

\def\repa{\raise4pt\hbox{$\square$}\mkern-14mu\raise-4pt\hbox{$\square$}}
\def\repab{\overline{\raise4pt\hbox{$\square$}\mkern-14mu\raise-4pt\hbox{$\square$}\mkern-1mu}}

\begin{document}

\title{
On the universality of entanglement entropy in quantum field theory
}

\author{Orr Sela}

\affiliation{
Laboratory for Theoretical Fundamental Physics, Institute of Physics, Ecole Polytechnique Federale de Lausanne (EPFL), CH-1015 Lausanne, Switzerland
}

\date{\today}

\begin{abstract}
The standard definition of entanglement entropy in quantum field theory requires regularizing the entangling surface and factorizing the Hilbert space, which in turn involves a choice of boundary conditions at the regularized surface. This raises two natural questions which are open at present. The first is under what conditions there is a choice of boundary conditions that leads to a unitary map between the original Hilbert space and the factorized one, which is a necessary condition for a well-defined entanglement entropy. The second concerns the situation where there are multiple such choices, in which case the question is whether they all yield the same entropy, or at least the same terms usually referred to as universal. Focusing on topological theories and using three-dimensional Chern-Simons theory as a concrete illustration, we show under specified assumptions that different such choices affect only the non-universal terms in the entropy, such that the remaining terms are universal in this sense as well, while we explicitly demonstrate that boundary conditions corresponding to non-unitary maps can change these terms. We then explain how a similar argument applies to two-dimensional conformal field theories and discuss a condition for universality in general theories. 
\end{abstract}

\maketitle

\section{Introduction}

Entanglement entropy has emerged as a central concept across diverse areas of physics, including quantum field theory, quantum gravity, condensed matter, and quantum information. However, its usual definition and computation require assigning Hilbert spaces to spatial subregions, even in common situations where the total Hilbert space does not admit a tensor-product decomposition.\footnote{Note that this absence of tensor-product decomposition is not special to the continuum, and also occurs in lattice gauge theories \cite{Donnelly:2011hn,Casini:2013rba,Radicevic:2014kqa}.} This, in turn, necessitates a choice of boundary condition at the entangling surface, with computations in the literature corresponding to different such choices (even if usually made implicitly), and thus raises the question of which boundary conditions lead to a well-defined and universal notion of entanglement entropy.

\begin{figure}[tp]
  \centering
  \includegraphics[width=0.5\textwidth]{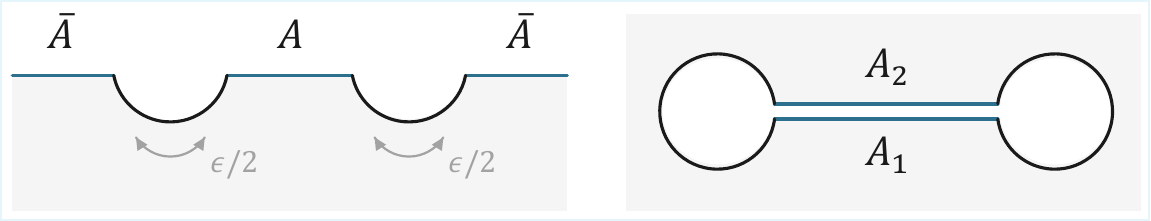}
  \caption{On the left side, the state in the factorized Hilbert space $H_{A}\otimes H_{\bar{A}}$ at $\tau=0$ (blue lines) corresponding to the original vacuum is depicted. It is given by the path integral on $\tau\leq0$ (shaded region) in the presence of the regularizing boundary (black semi-circles). On the right side, we present the path integral representation of the (unnormalized) reduced density matrix $\rho_A$, with boundary conditions $\psi_{A_{1}}$ and $\psi_{A_{2}}$ on $A_1$ and $A_2$ (respectively) corresponding to the matrix element $\langle\psi_{A_{2}}|\rho_{A}|\psi_{A_{1}}\rangle$. The figures are depicted for $d=2$, with the $d=3$ case obtained by rotating them around an axis passing through their middle.}
  \label{Map}
\end{figure}

In order to demonstrate this point explicitly, let us consider the vacuum state at $\tau=0$ of a quantum field theory on Euclidean $\mathbb{R}^{d}$ obtained by the path integral on $\tau\leq0$. Choosing for simplicity the entangling surface to be an $S^{d-2}$, it divides the $\tau=0$ slice to the interior of the sphere, which we denote by $A$, and to the complement exterior $\bar{A}$. Since the Hilbert space does not factorize into a part on $A$ times a part on $\bar{A}$, we excise a small tubular neighborhood of the entangling surface while imposing a boundary condition there,\footnote{A regularized entangling surface of this type is sometimes referred to in the literature as a brick wall \cite{tHooft:1984kcu} or a stretched horizon \cite{Susskind:1993if}.} thereby turning the $S^{d-2}$ into the codimension-one boundary $S^{d-2}\times S^1_\epsilon$ with $\epsilon$ being the circumference of $S^1_\epsilon$. This construction yields the desired Hilbert-space factorization, with the original vacuum mapped to the state obtained by the path integral on $\tau\leq0$ in the presence of the new boundary, see Fig. \ref{Map} (see also \cite{Ohmori:2014eia,Agia:2022srj} for more details on such a construction). At this point we can compute the reduced density matrix $\rho_{A}$ of the Hilbert space $H_A$ we assigned to $A$ using the boundary condition, by tracing over $H_{\bar{A}}$. The path integral representation of (the unnormalized) $\rho_{A}$ is given as usual by integrating over all the excised $\mathbb{R}^{d}$ except for a cut along $A$, where boundary conditions corresponding to matrix elements of $\rho_{A}$ are imposed, see Fig. \ref{Map}. The entanglement entropy can now in principle be computed using the standard formula 
\begin{equation}
S=-\textrm{Tr}_{H_A}\left(\rho_{A}\log\rho_{A}\right).
\label{S_def}
\end{equation}
However, in practice it is much easier to use the replica trick for this computation (see e.g. \cite{Calabrese:2009qy,Casini:2009sr,Rangamani:2016dms,Headrick:2019eth} for reviews), which is based on writing the trace in Eq. \eqref{S_def} in the form $\underset{n\rightarrow1}{\lim}\partial_{n}\textrm{Tr}_{H_{A}}\left(\rho_{A}^{n}\right)$ and using the path integral representation of $\textrm{Tr}_{H_A}(\rho_{A}^{n})$ as the partition function on an $n$-fold cover of the original manifold (here the excised $\mathbb{R}^{d}$). This $n$-fold cover manifold, known as the replica manifold, is obtained by gluing $n$ copies of the original manifold along the cut in $A$ in such a way that the fields at the lower part of the cut of each copy are identified (and summed over) with those at the upper part of the cut of the next copy. Notice that without regularizing the entangling surface by excising a small neighborhood of it the replica manifold would have a conical singularity along it, but in our case the manifold is completely smooth. Denoting the partition function on the replica manifold by $Z_n$, we then have $\textrm{Tr}_{H_A}(\rho_{A}^{n})=Z_n/Z_1^n$ where $Z_1$ is the partition function on the original regularized manifold, needed here to keep $\rho_{A}$ normalized. The entanglement entropy in \eqref{S_def} can therefore be computed from $Z_n$ as follows, 
\begin{equation}
S=\underset{n\rightarrow1}{\lim}\left(1-\partial_{n}\right)\log\left(Z_{n}\right),
\label{S_replica}
\end{equation}
which we will use below. Here $\epsilon$, the circumference of the transverse boundary circle at $n=1$, is held fixed in the replica computation and taken to be small only afterwards. 

In this shrinking limit, we also take the UV cutoff length to zero, keeping its ratio to $\epsilon$ fixed for simplicity.\footnote{The two regulators may instead be kept independent, and the universal part discussed below is extracted solely from the dependence on the region size at fixed regulators.} The entropy diverges in this limit, and for a scale-invariant theory its asymptotic form as $\epsilon/l\to0$, where $l$ denotes the size of the entangling sphere $S^{d-2}$ (say, the circumference of its great circle), is given by \cite{Liu:2012eea} 
\begin{equation}
\begin{aligned}
S= &\,\, a_{d-2}\left(l/\epsilon\right)^{d-2}+a_{d-4}\left(l/\epsilon\right)^{d-4}+\cdots \\ &+a_{1}\left(l/\epsilon\right)+(-1)^{\frac{d-1}{2}}s+\mathcal{O}(\epsilon)
\end{aligned}
\label{S_exp_odd}
\end{equation}
for odd dimensions $d$ and by 
\begin{equation}
\begin{aligned}
S= &\,\, a_{d-2}\left(l/\epsilon\right)^{d-2}+a_{d-4}\left(l/\epsilon\right)^{d-4}+\cdots \\
&+a_{2}\left(l/\epsilon\right)^{2}+(-1)^{\frac{d-2}{2}}s\log\left(l/\epsilon\right)+a_{0}+\mathcal{O}(\epsilon)
\end{aligned}
\label{S_exp_even}
\end{equation}
for even $d$. The coefficient $s$ is invariant under redefinitions of the regulator $\epsilon$, and for this reason it is referred to as the “universal part” of the entropy. It can be systematically obtained by acting on $S$ with a certain differential operator\footnote{For example, in 2d this operator is given by $l\partial_{l}$ and in 3d by $(l\partial_{l}-1)$, see \cite{Liu:2012eea} for the general form for any $d$.} and taking the $\epsilon\rightarrow0$ limit, in the context of which it is also sometimes called the renormalized entanglement entropy. For a renormalizable theory with mass scales $\mu$ that is not scale invariant, the coefficients in the $l$ expansions \eqref{S_exp_odd}-\eqref{S_exp_even} may depend on $\mu\epsilon$ but have a regular $\mu\epsilon\rightarrow0$ limit such that the divergences cannot be stronger than the ones above, see \cite{Liu:2012eea} for more details. In addition, $S$ can include terms such as $(\mu l)^{\alpha}$ and $(\mu l)^{\alpha}\log(\mu l)$ for $\alpha>0$ that would make the universal part (obtained using the same differential operator as in the scale-invariant case) $l$ dependent, allowing it to change along the RG flow. 

Notice that so far we did not refer at all to the boundary condition we have to impose at the regularized entangling surface, and it is natural to ask whether this choice can affect our discussion above. In fact, quite a dramatic observation is that not every boundary condition leads to a well-defined entanglement entropy. In some cases, the boundary conditions that we use to split the Hilbert space with result in non-unitary maps between the original Hilbert space and the factorized one, even at the $\epsilon\rightarrow0$ limit, such that the computation we described above does not capture the entanglement entropy of the original state. Indeed, acting with a non-unitary operator on a state will generally change the result of an entanglement computation. To understand better this requirement for a unitary map and what it implies for the allowed boundary conditions, we note that unitarity at the $\epsilon\rightarrow0$ limit is equivalent to the requirement that the regularized entangling surface shrinks to the identity defect (supported on the original entangling surface) at this limit. In the example we consider, it means that the boundary conditions should be such that the boundary $S^{d-2}\times S^1_\epsilon$ would result in the identity defect supported on $S^{d-2}$ at the $\epsilon\rightarrow0$ limit.\footnote{By this we mean that the leading term in the expansion of the boundary in terms of codimension-two defects at the small $\epsilon$ limit is given by the identity.} Indeed, such boundary conditions guarantee that any correlator computed with the factorized state in Fig. \ref{Map} will be equal at the $\epsilon\rightarrow0$ limit to the corresponding correlator computed with the original vacuum. In other words, this requirement, which is called \textit{shrinkability} in the literature \cite{Agia:2022srj,Donnelly:2016jet,Donnelly:2018ppr,Hung:2019bnq,Jafferis:2019wkd,Ball:2024hqe,Ball:2024xhf,Mertens:2022ujr,Wong:2025asd}, simply implies that taking the shrinking limit $\epsilon\rightarrow0$ removes the regularizing boundary and brings us back to the original setup.\footnote{Note that in practice this limit might be divergent and require counterterms, even when it leads to the identity defect. We will discuss it in detail below.} We will call a boundary condition that satisfies this condition "shrinkable", and will mainly be concerned with such boundary conditions as these are the ones that are required for a well-defined notion of entanglement entropy. Choosing instead a non-shrinkable boundary condition in the computation of the entropy would be equivalent to inserting extra codimension-two operators along the entangling surface, which will in general alter the result and will not represent an entanglement computation (this will be demonstrated explicitly in the next section). 

A general criterion for the existence of shrinkable boundary conditions in a generic theory is still lacking (although gravitational anomalies provide known obstructions to local Hilbert-space factorization~\cite{Hellerman:2021fla,Jensen:2017eof}). Nevertheless, even in theories where such boundaries are known to exist, a fundamental question that is crucial for the physical meaning one can associate to entanglement entropy is whether the universal terms depend on the specific choice of shrinkable boundary condition. As we will show below for topological theories, the universal terms can indeed change when non-shrinkable boundary conditions are employed, whereas under specified assumptions they are independent of the choice of shrinkable boundary condition and hence universal in this sense as well. We will then explain how a similar argument applies to two-dimensional conformal field theories (CFTs), before discussing the conditions for universality in general theories. In a forthcoming work~\cite{DS:toappear}, we will investigate shrinkability and universality in more general theories and in particular revisit, in light of these observations, entanglement entropy computations in the literature that yield different universal terms, such as those for four-dimensional Maxwell theory in \cite{Casini:2011kv,Solodukhin:2008dh,Donnelly:2014fua,Donnelly:2015hxa,Ball:2024hqe} versus \cite{Pretko:2015zva,Pretko:2018nsz} and \cite{Casini:2019nmu}.

\section{Entanglement entropy in Chern-Simons theory}

Let us begin by considering the simpler case of topological field theories, whose entanglement structure has been investigated extensively (see \cite{Kitaev:2005dm,LevinWen,Dong:2008ft,Grover:2011fa,Cano:2014pya,Wen:2016snr,Fliss:2017wop,Wong:2017pdm,Fliss:2023dze,Fliss:2023uiv,Sheffer:2025zyr} for an incomplete list). In order to be as concrete as possible, we will focus on three-dimensional $U(1)_k$ Chern-Simons theory with even $k>0$, although the various steps apply more generally. 

We consider the one dimensional Hilbert space of the theory on the sphere, and the state obtained by doing the path integral over the interior of the sphere \cite{Witten:1988hf}. The entangling surface, which is a line in this case, is chosen as the equator of the sphere, and as discussed in the previous section we regularize it and split the Hilbert space by removing a small tubular neighborhood of it (or a tube cut along its longer dimension in the case of the state on the sphere) while choosing a boundary condition there. 

Our first task in computing the entanglement entropy of this state is to find a shrinkable boundary condition. Specifically, since we are considering a circular entangling line, we are looking for a boundary condition for which a toric boundary which we introduce into a manifold $M$ would result in the identity line (possibly with a line counterterm) after shrinking its meridian cycle.\footnote{The meridian is the cycle of the torus which is contractible in its bulk, while the longitude is the cycle which is not.} To proceed, we first consider the manifold $M\setminus\mathsf{S}T^{2}$, where $\mathsf{S}T^{2}$ is the solid torus, with an unspecified boundary condition on the toric boundary such that the path integral over the manifold defines a state $|M\setminus\mathsf{S}T^{2}\rangle$ on this boundary.\footnote{By $M$ here we denote the manifold plus possible line insertions that do not cross the toric boundary.} To explore what this state looks like, we recall that the Hilbert space of the theory on the torus is $k$-dimensional with a basis of states $|\mathsf{S}T^{2};a\rangle$ given by the path integral on the solid torus (with unspecified boundary conditions) with Wilson line insertions $a$ along its non-contractible core \cite{Witten:1988hf,Elitzur:1989nr,Marino:2004uf}. These basis states are eigenstates of the meridian holonomy operator and are orthonormal. To see the first property, note that in the state $|\mathsf{S}T^{2};a\rangle$ a Wilson line insertion along the meridian direction of the boundary torus can be contracted through the bulk of the solid torus while generating a phase due to the line $a$ that it links. To see the orthonormality property, we note that the inner product between two such states is given by gluing using the identity map the two corresponding solid tori along their boundaries (where one of the tori is taken in an opposite orientation), which results in the manifold $S^2\times S^1$ with the two line insertions along the $S^1$ direction (and at different points on $S^2$), and computing the partition function. This then results in $\langle\mathsf{S}T^{2};a|\mathsf{S}T^{2};b\rangle=Z(S^{2}\times S^{1};\bar{a};b)=\delta_{ab}$ \cite{Witten:1988hf}, where $\bar{a}$ is the orientation reversal of $a$. 

We can now specify the state $|M\setminus\mathsf{S}T^{2}\rangle$ by expanding it in the basis $|\mathsf{S}T^{2};a\rangle$,  
\begin{equation}
|M\setminus\mathsf{S}T^{2}\rangle=\sum_{a}Z(M_{a})\,|\mathsf{S}T^{2};a\rangle
\end{equation}
where the product $Z(M_{a})=\langle\mathsf{S}T^{2};a|M\setminus\mathsf{S}T^{2}\rangle$ is obtained by gluing the solid torus with line $\bar{a}$ along the toric boundary of $M\setminus\mathsf{S}T^{2}$, i.e. it is the path integral on $M$ with the toric boundary removed (or filled) and with the Wilson line $\bar{a}$ inserted along its core. Choosing now a boundary condition $\mathcal{B}$ on the torus, we obtain for the partition functions 
\begin{equation}
Z_{\mathcal{B}}(M\setminus\mathsf{S}T^{2})=\sum_{a}Z(M_{a})Z_{\mathcal{B}}(\mathsf{S}T^{2};a)
\label{Z_b}
\end{equation}
where $Z_{\mathcal{B}}$ denotes the partition function with boundary condition $\mathcal{B}$, including integration over any additional boundary fields. Eq. \eqref{Z_b} shows us that the toric boundary $T_{\mathcal{B}}^{2}$ with the boundary condition $\mathcal{B}$ can be replaced by the sum over all the $k$ Wilson lines of the theory with coefficients $Z_{\mathcal{B}}(\mathsf{S}T^{2};a)$, that is,  
\begin{equation}
T_{\mathcal{B}}^{2}=\sum_{a}Z_{\mathcal{B}}(\mathsf{S}T^{2};a)\,\bar{a}\,.
\label{T_B}
\end{equation}
Therefore, in order to get a shrinkable boundary we need a boundary condition for which $Z_{\mathcal{B}}(\mathsf{S}T^{2};I)$ would be the leading term in the sum in Eq. \eqref{T_B} at the shrinking limit, where $I$ is the identity line, and that it would correspond to a line counterterm. This last requirement follows from the fact that line operators can always be rescaled by a counterterm given by the integral of a scalar built from local geometric quantities along the line, such that $Z_{\mathcal{B}}(\mathsf{S}T^{2};I)$ is allowed to be nontrivial (and even divergent at the shrinking limit) as long as it can be removed by such a counterterm. A simple boundary condition that satisfies these two properties is the familiar holomorphic boundary condition $A_{\bar z}\equiv A_m+iA_l=0$, where $A_m$ and $A_l$ are the boundary gauge-field components in the meridian and longitudinal directions, respectively. With this choice of boundary condition, which is very common in discussions of Chern-Simons theory on manifolds with boundaries, the boundary variation term $(ik/4\pi)\int_{T^{2}}A\wedge\delta A$ vanishes without the need to add boundary terms and the coefficients $Z_{A_{\bar z}=0}(\mathsf{S}T^{2};a)$ in \eqref{T_B} are simply the $\mathfrak{u}(1)_{k}$ characters $\chi_{n_{a}}(\tau)$ corresponding to the charges $n_{a}\in\mathbb{Z}_{k}$ of the lines $a$ \cite{Elitzur:1989nr,Bos:1989kn,Axelrod:1989xt,Porrati:2019knx,Porrati:2021sdc}. Considering a rectangular toric boundary of longitudinal length $l$ and meridian length $\epsilon$, i.e. $\tau=il/\epsilon$, these characters can be viewed as the partition functions of the edge-mode Hilbert spaces on the meridian circle of size $\epsilon$ at inverse temperature $\beta=l$, and have the following asymptotic behavior as $\epsilon\rightarrow0$, 
\begin{equation}
\label{Chi_asym}
\chi_{n_a}(\tau)\simeq
\bigl(1+\delta_{\widetilde{n}_a,k/2}\bigr)
\exp\!\left[
-\frac{2\pi l}{\epsilon}
\left(\frac{\widetilde{n}_a^2}{2k}-\frac{1}{24}\right)
\right]
\end{equation}
where $\widetilde{n}_a$ is the integer representative of $n_a$
in $(-k/2,k/2]$. We then have at the leading shrinking limit\footnote{See also Ref.~\cite{Sheffer:2025zyr} for a related gluing construction using vacuum dominance.}  
\begin{equation}
T_{A_{\bar z}=0}^{2}=\sum_{a}\chi_{n_{a}}(\tau)\,\bar{a}\simeq\exp\left(\frac{\pi l}{12\epsilon}\right)I
\label{T_Az}
\end{equation}
and we indeed see that the boundary shrinks to the identity line along a loop of length $l$, with a factor corresponding to the cosmological constant counterterm on $I$ (given explicitly by $\exp\left(\frac{\pi}{12\epsilon}\oint ds\right)$). As we discussed, this factor can be rescaled away and we conclude that $T_{A_{\bar z}=0}^{2}$ is shrinkable. 

Now that we have found an example of a shrinkable boundary, we note that other such boundaries are possible, including ones that couple the gauge field to a nontrivial boundary theory. Since they are all shrinkable, they yield the identity line upon shrinking, but the accompanying counterterms may differ. As a simple example, let us consider the shrinkable boundary $T_{A_{\bar z}=0}^{2}$ and add to it a 2d CFT $\mathcal{T}$ with central charge $c$ which is decoupled from the 3d bulk theory. Then, at the shrinking limit $\epsilon\rightarrow0$ (where $\tau=il/\epsilon$ as in \eqref{Chi_asym}), which is equivalent to the low-temperature limit of the 2d CFT on the torus, we have inside any correlator of the bulk theory
\begin{equation}
T_{A_{\bar z}=0;\mathcal{T}}^{2}\simeq\exp\left(\frac{\pi l}{12\epsilon}\left(1+2c\right)\right)I
\label{T_Am_decoupledCFT}
\end{equation}
where $T_{A_{\bar z}=0;\mathcal{T}}^{2}$ denotes the toric boundary with the decoupled CFT. Since the exponential factor on the right hand side is again a cosmological constant counterterm on $I$, this boundary is also shrinkable but clearly involves a different counterterm than in \eqref{T_Az}. 

It will be useful at this point to contrast the above examples of shrinkable boundaries with an example of a non-shrinkable one. For this purpose, let us momentarily set aside the $U(1)_k$ theory under consideration and instead examine a $U(1)_k\times U(1)_{-k}$ theory. Denoting its two gauge fields by $A$ and $B$, we choose the topological boundary condition $(A-B)_{T^2}=0$ which is known to result in the sum of the Wilson lines that are condensed at the boundary (known also as the "algebra object" associated with this boundary), 
\begin{equation}
T_{A=B}^{2}=\sum_{n\in\mathbb{Z}_{k}}\exp\left[in\oint(A-B)\right],
\label{T_topo}
\end{equation}
see \cite{Kaidi:2021gbs} for a recent detailed discussion.\footnote{Eq. \eqref{T_topo} can also be obtained from \eqref{T_B} by noticing that the solid-torus partition function with this boundary condition equals to one for the $A-B$ core lines appearing in (11), and vanishes for all other line types.} This boundary is clearly not shrinkable, and results in an (unnormalized) projector to the zero-charge sector of one of the $\mathbb{Z}_{k}$ 1-form symmetries of the theory \cite{Gaiotto:2014kfa}. A sharp consequence of this lack of shrinkability can be observed when trying to use this boundary to factorize Hilbert spaces. As an example, for $k>2$, let us consider the one dimensional Hilbert space of the theory on the sphere with two marked points (or punctures) corresponding to a bulk Wilson line $\exp[im\int(A+B)]$ (note the plus sign) with $2m\not=0\,\textrm{mod}\,k$ ending on the boundary sphere at these two points. To split this Hilbert space, we remove as before a small tubular neighborhood of the equator while imposing the topological boundary condition $A=B$ there. Moreover, we take the two marked points to be from the opposite sides of the regularized equator. Then, the original state obtained by performing the path integral over the solid sphere (with the $A+B$ line and without the tubular boundary) is mapped by this construction to the state in the factorized Hilbert space corresponding to the path integral on the solid sphere in the presence of the tubular boundary. We can now compute the norm of this factorized state by gluing the regularized solid sphere to another copy of it (with an opposite orientation) along their non-tubular boundaries, resulting in $S^3$ with a thin toric boundary along one of its great circles and with an $A+B$ Wilson loop circling it. However, since according to Eq. \eqref{T_topo} the toric boundary can be replaced with an operator that projects to the zero-charge sector of the symmetry that acts on the $A+B$ lines, and since this operator links the $A+B$ Wilson loop, the path integral on this manifold vanishes. We therefore obtain that the initial state, whose norm is $k^{-1/2}$,\footnote{To see this, note that the product of the initial state with itself results in the path integral on $S^3$ with a Wilson loop $\exp[im\oint(A+B)]$ inside, which is equal to $k^{-1}$ \cite{Witten:1988hf}.} is mapped to a null state such that the map based on the topological boundary condition is clearly not unitary. 

\begin{figure}[tp]
  \centering
  \includegraphics[width=0.5\textwidth]{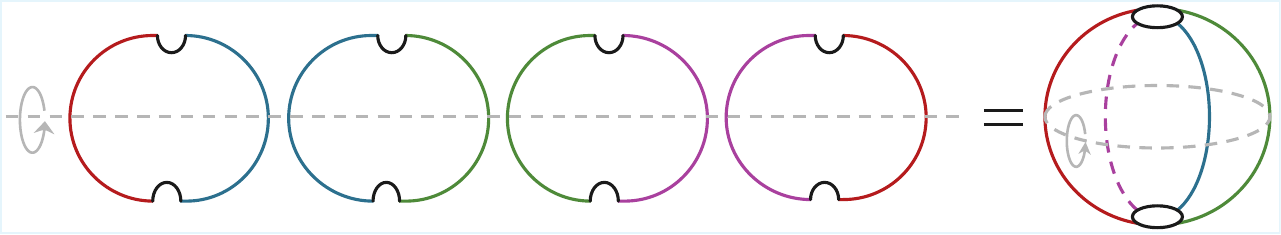}
  \caption{Construction of the replica manifold $M_n$, here depicted for $n=2$. Each regularized solid sphere is represented by a regularized disk rotated along an axis passing through its center (gray dashed line), and the gluing is done along same-color hemispheres.}
  \label{Replica}
\end{figure}

Returning to the $U(1)_k$ theory, we are now in a position to compute the entanglement entropy of the state on the sphere with the entangling line being the equator, and analyze how different shrinkable boundary conditions affect the result. In order to do this, we glue two copies of the regularized solid sphere along one hemisphere to obtain the (unnormalized) reduced density matrix, and then construct the replica manifold $M_n$ by gluing $n$ such density matrices appropriately. Overall, we use $2n$ copies of the regularized solid sphere, see Fig. \ref{Replica} for a depiction of this construction following Ref. \cite{Dong:2008ft}. 

A key feature of $M_n$ is that its thin toric boundary (corresponding to the two black circles on the right hand side of Fig. \ref{Replica}) regularizes a would-be conical singularity. Indeed, notice that if $\epsilon$ is the meridian length of the toric boundary used in the reduced density matrix geometry (or $M_1$, obtained by gluing two copies of the regularized solid sphere), then $n\epsilon$ is the corresponding length in $M_n$ even though the radius of this meridian circle (in the conical geometry obtained by filling the toric boundary) is still $\epsilon/2\pi$. In other words, there is an excess angle of $2\pi(n-1)$ in the meridian direction. However, since the theory living in the bulk of $M_n$ is topological, we can freely deform this bulk (i.e. its metric) while keeping the induced metric on the toric boundary fixed, in such a way that now after shrinking the boundary there will be no conical singularity (this is analogous to flattening a regularized cone into a holed plane). In particular, by such a deformation we can turn $M_n$ into $S^{3}\setminus\mathsf{S}T^{2}$ without changing the induced metric on the toric boundary, which for $U(1)_k$ Chern-Simons theory is necessarily not topological. See Appendix~\ref{app:flattening} for the explicit deformation and its effect on boundary couplings involving extrinsic or ambient curvature. 

At this point we can consider the leading behavior in the shrinking limit $\epsilon\to0$. Since we are considering shrinkable boundary conditions, the boundary reduces at leading order to the identity line, possibly accompanied by line-defect counterterms, leaving $S^3$ with these counterterms. Focusing on the cosmological constant (or perimeter) line counterterm $\exp\left(\alpha l/\epsilon\right)$ (with $\alpha$ taken to be independent of $n$ for simplicity), which is the only one that can yield a divergent term in the entropy, the path integral on the replica manifold is given by 
\begin{equation}
Z(M_{n})=Z(S^{3})\exp\left(\frac{\alpha l}{n\epsilon}\right)=\frac{1}{\sqrt{k}}\exp\left(\frac{\alpha l}{n\epsilon}\right)
\label{ZMn}
\end{equation}
up to subleading terms in $\epsilon$. Note that $n\epsilon$ appears in \eqref{ZMn} instead of $\epsilon$ since as we discussed before, the meridian length of the toric boundary of $M_n$ is $n\epsilon$. We can now use Eq. \eqref{S_replica} to find the entanglement entropy, 
\begin{equation}
S=2\alpha\left(l/\epsilon\right)-\frac{1}{2}\log(k)
\label{EE_CS}
\end{equation}
with subleading terms in \eqref{ZMn} corresponding to subleading terms in \eqref{EE_CS}.\footnote{Here we use the leading shrinking behavior on the replica manifold before evaluating the replica derivative at $n=1$, reversing the order prescribed by Eq.~\eqref{S_replica}. In Appendix~\ref{app:flattening} we show that under the assumptions stated there, this changes the entropy only by terms that vanish as $\epsilon\to0$.} Note that we did not make any specific choice of shrinkable boundary condition in this computation, with different such choices affecting only $\alpha$ (since they may result in different counterterms) and the subleading terms in \eqref{EE_CS}. For example, the two shrinkable boundaries we discussed before, $T_{A_{\bar z}=0}^{2}$ and $T_{A_{\bar z}=0;\mathcal{T}}^{2}$, correspond to $\alpha=\pi/12$ and $\alpha=\pi(1+2c)/12$, respectively. We therefore see that the universal constant term in \eqref{EE_CS}, known in general for topological field theories as the \textit{topological entanglement entropy} \cite{Kitaev:2005dm,LevinWen}, is also universal in the sense that it is independent of the choice of shrinkable boundary condition used in the computation (and not only universal under changing $\epsilon$ or the scheme in a computation with a given boundary condition). In particular, different counterterms arising from using different boundary conditions cannot affect this value. Looking at this in the opposite direction, changing this constant would require adding a multiplicative factor to the identity defect (plus the counterterm) obtained after the shrinking in the path integral in \eqref{ZMn}, which would in turn mean that the boundary used in the computation is in fact not shrinkable. As an example, let us consider again the $U(1)_k\times U(1)_{-k}$ theory, whose topological entanglement entropy as computed using the procedure above is twice the value in \eqref{EE_CS}, which itself is the familiar result for $U(1)_k$ Chern-Simons theory \cite{Kitaev:2005dm,LevinWen}. However, using instead the boundary $T_{A=B}^{2}$ in the computation would result in a vanishing topological entanglement entropy,\footnote{To see this, note that the sum over lines in \eqref{T_topo} would give a factor of $k$ since the lines are supported on contractible loops and have a quantum dimension which is equal to one. Together with the factor of $k^{-1}$ coming from $Z(S^{3})$, the overall constant factor in $Z(M_n)$ would be one, corresponding to a vanishing topological entanglement entropy.} reflecting the non-shrinkability of the boundary condition. 

Notice that our universality argument relies only on shrinkability and bulk topological invariance, and therefore applies to any topological field theory admitting shrinkable boundaries satisfying the assumptions in Appendix~\ref{app:flattening}. In addition, it also extends to any state prepared by insertions that remain separated from the regulating boundary. 

\section{Universality in general theories}

A crucial step in the universality argument of the previous section was the deformation of the bulk metric near the regulating boundary of the replica manifold, keeping its induced metric fixed, so that the boundary could be filled without a conical singularity. Bulk topological invariance allowed this deformation without changing the path integral, and hence enabled us to apply ordinary shrinkability in a smooth geometry. However, such deformations generally change the path integral in a non-topological theory, so shrinkability on smooth backgrounds may not suffice for the replica construction. Indeed, additional requirements might be needed for a well-defined entanglement entropy, possibly involving an extension of shrinkability to replica geometries.\footnote{Another requirement, assumed throughout, is locality of the regulating boundary conditions, needed for replica gluing to represent powers of a density matrix~\cite{Agia:2022srj}.}

A general understanding of these requirements, and of when boundary conditions satisfying them exist, remains open. In fact, even the conditions for the existence of ordinary shrinkable boundaries are not established for general theories, and we defer these questions to future work. Here, instead, we will focus on when different local shrinkable boundary conditions yield the same universal terms in the entropy, without claiming to establish their full admissibility. 

Let us first consider two-dimensional CFTs, where conformal symmetry allows a universality argument analogous to that for topological theories, and then examine why the same reasoning does not directly extend to higher-dimensional CFTs. For concreteness, we consider the vacuum of a two-dimensional CFT with $c_L=c_R=c$ on the plane, with the entangling region an interval of length $l$. We regularize each endpoint by excising a disk of circumference $\epsilon$ and imposing a local shrinkable boundary condition, and construct $M_n$ by replica gluing. Universality in this setting was also argued in Ref.~\cite{Agia:2022srj} using angular quantization and conformal maps that also transform the boundary conditions. Here we use a local Weyl rescaling that preserves the induced metric on the regulating boundaries.\footnote{This rescaling can be confined to sufficiently small neighborhoods of the regulating boundaries, avoiding bulk operator insertions kept at fixed nonzero distance from the entangling points. The argument therefore also applies in their presence.}

Near either regulating circle, the metric of $M_n$ is $ds^2=dr^2+n^2r^2d\theta^2$, where $0\leq\theta<2\pi$ and $r\geq\epsilon/2\pi$. At finite $\epsilon$, we perform the Weyl rescaling $ds^2\rightarrow e^{-2\phi(r)}ds^2$ and introduce the radial coordinate 
\begin{equation}
\mathcal{R}=nr\left(\frac{\epsilon}{2\pi r}\right)^{1-\frac{1}{n}}\,,
\;\,
\phi(r)=g(r)\left(1-\frac{1}{n}\right)
\log\left(\frac{2\pi r}{\epsilon}\right).
\label{Trans_2d}
\end{equation}
Here $g(r)$ is a smooth function equal to one for $r\in[\epsilon/2\pi,\epsilon/2\pi+\delta_1]$ and zero for $r\geq\epsilon/2\pi+\delta_2$, with $0<\delta_1<\delta_2$ chosen so that the $\delta_2$-neighborhoods of the two regulating boundaries are disjoint. Where $g(r)=1$, the transformed metric is the flat $d\tilde{s}^{\,2}=d\mathcal{R}^2+\mathcal{R}^2d\theta^2$, and the regulating circle lies at $\mathcal{R}=n\epsilon/2\pi$. Since $\phi(\epsilon/2\pi)=0$, the Weyl rescaling acts trivially on the induced metric on the regulating boundary, while also being trivial outside these $\delta_2$-neighborhoods.\footnote{For boundary interactions $\lambda_iK^{p_i}\mathcal O_i$ involving the extrinsic curvature $K$, the rescaling in Eq.~\eqref{Trans_2d} sends $K\to K/n$, so we preserve these interactions by shifting $\lambda_i\to n^{p_i}\lambda_i$ and assume ordinary shrinkability persists for the shifted couplings near $n=1$ (meanwhile, ambient curvature couplings vanish near the regulating boundaries before and after the rescaling). See Appendix~\ref{app:flattening} for the analogous discussion in topological theories.}

We can therefore apply ordinary shrinkability in the locally flat geometry, replacing each regulating boundary at leading order by the identity operator, up to local counterterms that amount to multiplicative normalization factors at the entangling points. Note that the Weyl rescaling changes the partition function by an anomaly factor, however it is independent of the boundary condition and therefore cancels when two such partition functions are compared.\footnote{More explicitly, the neighborhood of each regulating circle contributes a Weyl factor $\exp\left(\frac{nc}{12}\int_{\epsilon/2\pi}^{\epsilon/2\pi+\delta_{2}}(\partial_{r}\phi)^{2}rdr\right)$, which clearly does not depend on the chosen boundary condition. The boundary term, with exponent proportional to $\int_{\partial M_n}ds\,K\phi$, vanishes because $\phi=0$ on the regulating boundaries.} The remaining local counterterms do not depend on the interval length and hence cannot change the universal logarithmic dependence appearing in the entropy. In other words, different shrinkable boundary conditions, which at most can result in different counterterms, necessarily give rise to the same universal term. For the vacuum state and the interval we consider here, this universal term is the familiar $(c/3)\log(l/\epsilon)$, while the non-universal additive constant depends on the boundary condition \cite{Agia:2022srj} (see also \cite{Michel:2016fex}). 

Let us comment that the preceding universality argument assumes that retaining the leading small-$\epsilon$ behavior of the replica partition function before evaluating the replica derivative at $n=1$ leaves the universal logarithmic term unchanged (as is also implicitly assumed in Ref.~\cite{Agia:2022srj}). Whereas this step was justified for topological theories under the assumptions in Appendix~\ref{app:flattening}, here we impose it as an additional assumption on the local shrinkable boundary conditions to which our universality statement applies.

Turning to higher-dimensional CFTs, the obstruction to applying the same argument can already be seen from the leading metric near a smooth entangling surface $\Sigma$, $ds^2\simeq dr^2+n^2r^2d\theta^2+h_{ij}(y)dy^i dy^j$, where $h_{ij}$ is the induced metric on $\Sigma$. Applying the Weyl rescaling and coordinate transformation in Eq.~\eqref{Trans_2d}, we obtain in the region where $g(r)=1$ the metric 
\begin{equation}
d\tilde{s}^{\,2}\simeq
d\mathcal{R}^2+\mathcal{R}^2d\theta^2
+\left(\frac{n\epsilon}{2\pi\mathcal{R}}\right)^{2(n-1)}
h_{ij}(y)dy^i dy^j.
\label{Trans_higher_d}
\end{equation}
Although the transverse metric becomes flat and the induced boundary metric remains unchanged, the longitudinal warp factor generates curvature: even when $M_1$ is flat and $\Sigma$ is planar, the Ricci scalar of the transformed metric is $\widetilde R=-(d-1)(d-2)(n-1)^2/\mathcal{R}^2$, which diverges at the regulating boundary $\mathcal{R}=n\epsilon/(2\pi)$ as $\epsilon\to0$ for fixed $n\ne1$. Thus this transformation does not yield a smooth shrinking limit, and ordinary shrinkability alone does not justify the same universality argument. 

Nevertheless, whenever the leading small-$\epsilon$ behavior can be used before evaluating the replica derivative at $n=1$ without changing the universal terms, two shrinkable boundary conditions would yield the same universal terms in the entropy when their leading replica partition functions are related near $n=1$ by a counterterm rescaling along $\Sigma$, with coefficients allowed to depend smoothly on $n$.\footnote{Here $Z_{\mathrm{ct}}=\exp[\int_\Sigma d^{d-2}y\,\sqrt h\,\sum_i c_i(n,\epsilon)\mathcal I_i]$, where $\mathcal I_i$ are local geometric scalars constructed from the induced metric, extrinsic curvature and ambient curvature.} Ordinary shrinkability ensures such a relation on smooth backgrounds, but whether it extends to the boundaries regulating conical singularities remains to be understood in general. 

To formulate an exact universality condition directly on the smooth $n=1$ manifold without interchanging the replica derivative and the shrinking limit, we next apply Eq.~\eqref{S_replica} at finite $\epsilon$. Denoting the unnormalized reduced density matrix for the regulating boundary condition $\mathcal B$ by $\widehat\rho_{\mathcal B,\epsilon}$ and defining $\mathcal K_{\mathcal B,\epsilon}=-\log\widehat\rho_{\mathcal B,\epsilon}$, we have $Z_{n,\mathcal B}=\operatorname{Tr}\widehat\rho_{\mathcal B,\epsilon}^{\,n}$ and hence
\begin{equation}
S_{\mathcal B}(\epsilon)
=\log Z_{1,\mathcal B}(\epsilon)
+\left\langle\mathcal K_{\mathcal B,\epsilon}\right\rangle_{1,\mathcal B}.
\label{S_insertion}
\end{equation}
Here $Z_{1,\mathcal B}$ is the partition function on the regulated $M_1$, while the second term is the normalized expectation value obtained by inserting the operator $\mathcal K_{\mathcal B,\epsilon}$ in the same path integral, with $\epsilon$ taken small only afterwards. When changing $n$ near $1$ can be implemented solely by varying the metric, this expectation value, obtained from the $n$-derivative in Eq.~\eqref{S_replica}, can be expressed through a weighted integral of the full stress tensor (which includes all the contributions supported on the regulating boundary). We now notice that ordinary shrinkability allows us to remove the regulating boundary from $Z_{1,\mathcal B}$ as $\epsilon\to0$, changing the first term only by counterterms, so the remaining condition for universality is that the small-$\epsilon$ behavior of the second term agrees in its universal term for different shrinkable boundary conditions. This, in turn, does not follow directly from shrinkability, since the insertion extends to the regulating boundary rather than remaining at a fixed distance from it. Thus, given ordinary shrinkability, universality is determined by the shrinking behavior of the regulating boundary in the presence of the $\mathcal K_{\mathcal B,\epsilon}$ insertion. 

As an example, let us consider the vacuum of a four-dimensional CFT with a spherical entangling surface of radius $R$ in flat space. The causal domain of the enclosed ball is conformally equivalent to a Rindler wedge~\cite{Casini:2011kv}, and we assume that the theory admits local shrinkable boundary conditions in the spherical setup for which the regulating boundary and its boundary condition become independent of Rindler time under this map (i.e. they preserve the spherical modular flow). Then, in the resulting Euclidean Rindler setup, this time independence means that replication only multiplies the duration of angular evolution by $n$, such that differentiating with respect to $n$
inserts $2\pi$ times the generator of angular evolution, including its boundary contributions. Mapping back to the spherical setup then gives (see \cite{Casini:2011kv} for the explicit form of the map) 
\begin{equation}\label{K_T}
\mathcal K_{\mathcal B,\epsilon}
=-2\pi\int_{A_\epsilon}d^3x\,
\frac{R^2-r^2}{2R}\,
T_{\tau\tau}^{E,\mathrm{full}}(0,\mathbf x)
\,,
\end{equation}
where $A_\epsilon$ is the regulated ball on the $\tau=0$ slice, and the full Euclidean stress tensor $T_{\mu\nu}^{E,\mathrm{full}}$ is obtained by varying the complete bulk and boundary action for boundary condition $\mathcal B$ at finite $\epsilon$, such that the integral includes all the contributions supported on $\partial A_\epsilon$. Considering now the entropy given in \eqref{S_insertion}, shrinkability implies that removing from $\log Z_{1,\mathcal B}$ the regulating boundary as $\epsilon\to0$ results in $\log Z(\mathbb R^4)$ along with counterterms supported on the entangling sphere, which do not contain a universal $\log(R/\epsilon)$ term. As a result, such a term must originate from the small-$\epsilon$ behavior of the boundary with the insertion \eqref{K_T}, and universality reduces to matching its coefficient between different such shrinkable boundary conditions. This, in turn, might require information beyond ordinary shrinkability. 

Explicit analyses of shrinkability and universality in various theories, including four-dimensional CFTs such as Maxwell theory, will appear in a forthcoming work~\cite{DS:toappear}.

\section{Acknowledgments} I would like to thank Thomas Dumitrescu for introducing me to the questions addressed in this note, for comments on the manuscript and for collaboration on a related project. This work was supported by the Simons Foundation grant 994310 (Simons Collaboration on Confinement and QCD Strings) and by a grant of access to OpenAI models through the ChatGPT for Academic Researchers program. The AI models were used to a limited extent, primarily to cross-check computations.

\bibliographystyle{JHEP}
\bibliography{refs}

\onecolumngrid
\clearpage
\appendix
\setcounter{secnumdepth}{1}
\section{Smoothing the replica geometry and the shrinking limit}
\label{app:flattening}

A key step in the universality argument for topological field theories was to deform the metric of the replica manifold $M_n$ near the regulating boundary, keeping its induced metric fixed, so that shrinking the boundary would leave no conical singularity. In this appendix we describe this deformation and its implications
for boundary conditions that depend on geometric
data beyond the induced metric, such as extrinsic
or ambient curvature. Moreover, the entropy computation used the leading $\epsilon\to0$ behavior on $M_n$ before evaluating the replica derivative at $n=1$, reversing the order prescribed by Eq.~\eqref{S_replica}. We will explain why this reproduces the entropy up to terms vanishing as $\epsilon\to0$ under the assumptions specified below. 

Taking the three-ball used to prepare the state
on $S^2$ to have the metric of a hemisphere
of a round $S^3$ of radius $l/(2\pi)$, the gluing
construction described in the main text gives the
regularized replica manifold $M_n$ with metric
\begin{equation}
ds_n^2=
\frac{dr^2}{1-(2\pi r/l)^2}
+n^2r^2d\theta^2
+\left[1-(2\pi r/l)^2\right]dy^2.
\label{app:round_metric}
\end{equation}
Here $l$ is the length of the entangling great circle, $y\sim y+l$ runs along it, $\theta\sim\theta+2\pi$ is the transverse angular coordinate, and $r$ is a transverse radial coordinate chosen so that the meridian circumference is $2\pi nr$. The regulating boundary therefore lies exactly at $r=\epsilon/(2\pi)$, where its induced metric describes a flat rectangular torus with meridian
length $n\epsilon$ and longitude length
$\sqrt{l^2-\epsilon^2}$. For $\epsilon\ll l$, the
longitude approaches $l$, while in the near-boundary region $r\ll l$ the metric reduces to
$ds_n^2\simeq dr^2+n^2r^2d\theta^2+dy^2$,
with corrections of order $(r/l)^2$
in the radial and longitudinal components. We can now deform this geometry near the regulating boundary by multiplying $g_{rr}$ by $n^2$, leaving its induced metric unchanged. With $R=nr$, the near-boundary metric then becomes $d\tilde s^2\simeq dR^2+R^2d\theta^2+dy^2$, which admits a smooth extension through $R=0$ such that shrinking the boundary no longer produces a conical singularity. This deformation can be smoothly interpolated to the original metric away from the boundary, and the resulting smooth filling has the topology of $S^3$, whose partition function is simply $Z(S^3)$ by topological invariance. 

The deformation we just described preserves the induced metric on the boundary, so boundary interactions whose geometric dependence is entirely through this metric (as is typically the case) remain unchanged. However, interactions involving extrinsic or ambient curvature require additional care, since the deformation rescales the extrinsic-curvature tensor by $1/n$, and, for the round geometry in Eq.~\eqref{app:round_metric}, the ambient Riemann tensor in an orthonormal frame by $1/n^2$. Consequently, an interaction with coupling $\lambda_i$ involving a boundary operator and contractions of $p_i$ powers of extrinsic curvature and $q_i$ powers of ambient curvature, is reproduced in the deformed geometry by replacing its coupling with $\lambda_i(n)=n^{p_i+2q_i}\lambda_i$, while interactions involving covariant derivatives of ambient curvature vanish in both geometries. In other words, we compensate for the changes in extrinsic and ambient curvature by adjusting the couplings $\lambda_i$, and assume that ordinary shrinkability continues to hold on the resulting smooth geometries for the shifted couplings $\lambda_i(n)$ with $n$ near $1$. This requires shrinkability to persist under these small changes of the couplings and of the smooth geometry, and could fail, for example, if shrinkability holds only at the original coupling values. 

Let us next examine why retaining the leading shrinking behavior on the replica manifold before evaluating the replica derivative at $n=1$, reversing the order prescribed by Eq.~\eqref{S_replica}, leaves the universal constant in the entropy unchanged. After the smoothing deformation, bulk topological invariance gives the same contribution $\log Z(S^3)$ for all $n$, while the remaining $n$ dependence comes from the regulating boundary and its interactions. In the geometry considered above, this dependence enters through the meridian length $n\epsilon$, the transformed couplings $\lambda_i(n)=n^{p_i+2q_i}\lambda_i$, and the ambient Ricci scalar $\kappa/n^2$, where $\kappa=24\pi^2/l^2$ (while the extrinsic curvature is determined by these parameters). We therefore write 
\begin{equation}
\log Z_n=\log Z(S^3)+C(n,\epsilon)+F\!\left(n\epsilon,\{n^{p_i+2q_i}\lambda_i\},\frac{\kappa}{n^2}\right),
\end{equation}
where $C$ collects the counterterms obtained after shrinking and $F$ is the remainder, with its dependence on $l$ and other unchanged parameters suppressed. Note that the curvature dependence of $F$ arises through the boundary geometry and interactions, despite the topological nature of the bulk. Following Eq.~\eqref{S_replica} at finite $\epsilon$, the contribution of this remainder to the entropy is then 
\begin{equation}
\delta S=\left[1-\epsilon\partial_\epsilon-\sum_i(p_i+2q_i)\lambda_i\partial_{\lambda_i}+2\kappa\partial_\kappa\right]F(\epsilon,\{\lambda_i\},\kappa).
\end{equation}
Now let us consider small $\epsilon$. Shrinkability implies that $F(\epsilon,\{\lambda_i\},\kappa)\to0$ as $\epsilon\to0$, so the first term vanishes. Moreover, if $\epsilon\partial_\epsilon F$ has a finite limit, this limit must also vanish since a nonzero value would force logarithmic behavior in $\epsilon$, contradicting shrinkability. The remaining terms describe differentiation along the family of adjusted couplings $\lambda_i(n)$ and ambient curvatures $\kappa/n^2$ produced by the smoothing, with the meridian length and $l$ held fixed. The shrinkability assumed above along this family then implies $F(\epsilon,\{\lambda_i(n)\},\kappa/n^2)\to0$ for every $n$ near $1$, and its derivative along the family also tends to zero provided that this derivative converges uniformly to a continuous limit in a fixed neighborhood of $n=1$. Indeed, integrating the derivative between $1$ and any nearby $n$ then gives the difference between two vanishing remainders, forcing the limiting derivative to vanish throughout that neighborhood.

We have thus assumed that $\epsilon\partial_\epsilon F$ has a well-defined finite limit as $\epsilon\to0$ and that the derivative along the family of couplings and smooth geometries described above converges uniformly. Under these assumptions, $\delta S\to0$, so retaining the leading shrinking behavior before evaluating the replica derivative reproduces the entropy up to terms vanishing as $\epsilon\to0$, and in particular leaves its universal constant unchanged.

\end{document}